\documentclass{article}

\usepackage{PRIMEarxiv}

\usepackage[utf8]{inputenc} 
\usepackage[T1]{fontenc}    
\usepackage{subcaption}
\usepackage{hyperref}       
\usepackage{url}            
\usepackage{booktabs}       
\usepackage{amsfonts}       
\usepackage{amsmath}
\usepackage{amsthm}
\usepackage{amssymb}
\usepackage{tikz-cd}
\usepackage{adjustbox}
\usepackage{nicefrac}       
\usepackage{microtype}      
\usepackage{lipsum}
\usepackage{tikz}
\usepackage{fancyhdr}       
\usepackage{graphicx}       
\usepackage{float}
\graphicspath{{media/}}     

\title{Quantum generative adversarial networks for gluon-initiated jets generation
}

\author{
Rey Guadarrama \\
Fac. de Cs. Físico-Matemáticas
BUAP, \\
Puebla, Mexico 72570 \\
\texttt{luis.vargazg@alumno.buap.mx} \\  
\And
Sergei Gleyzer \\
Dep.~Physics \& Astronomy\\
University of Alabama\\
Tuscaloosa, AL 35487 \\
\texttt{} \\
\And
Mariia Baidachna \\
University of Glasgow \\ 
Glasgow G12 8QQ, Reino Unido\\ 
\texttt{}
\And
Kyoungchul Kong \\
Dep.~Physics \& Astronomy\\
University of Kansas\\
Lawrence, KS 66045 \\
\texttt{} \\
\And
Konstantin T. Matchev \\
Physics Department, \\
University of Florida \\ 
Gainesville, FL 32611\\ 
\texttt{} \\
\And
Katia Matcheva \\
Physics Department, \\
University of Florida \\ 
Gainesville, FL 32611\\   
\texttt{} \\
\And
Isabel Pedraza \\
CIIEC, \\
BUAP. \\ 
Puebla, Mexico 72570\\ 
\texttt{}
\And
Gopal Ramesh Dahale \\
École Polytechnique Fédérale de Lausanne\\
University of Alabama\\
CH-1015 Lausanne \\
\texttt{} \\
\And
Haydee Hernández-Arellano \\
CIIEC, \\
BUAP. \\ 
Puebla, Mexico 72570\\ 
\texttt{}
}

\begin{document}
\maketitle

\begin{abstract}
Quantum computing has the potential to offer significant advantages over classical computing, making it a promising avenue for exploring alternative methods in High Energy Physics (HEP) simulations. This work presents the implementation of a Quantum Generative Adversarial Network (qGAN) to simultaneously generate gluon-initiated jet images for both ECAL and HCAL detector channels, a task crucial for high-energy physics simulations at the Large Hadron Collider (LHC). The results demonstrate high fidelity in replicating energy deposit patterns and preserving the implicit training data features. This study marks the first step toward generating multi-channel pictures and quark-initiated jet images using quantum computing.
\end{abstract}

\section{Introduction}
Monte Carlo simulations have long been essential in computational physics, providing a robust framework for modeling complex systems and stochastic processes \cite{metropolis}, \cite{Carlson} In HEP, Monte Carlo methods are necessary for simulating particle interactions and understanding fundamental forces \cite{Colin}. However, the computational intensity of these simulations often limits their scalability and efficiency, especially when dealing with large datasets or complex models. In recent years, Generative Adversarial Networks (GANs) have emerged as powerful tools for generating realistic data samples \cite{goodfellow}, \cite{goodfellow2016tutorial}. Although GANs have achieved wide success in HEP \cite{Musella_2018}, \cite{Di_Sipio_2019}, a variety of problems during training in practice, namely vanishing gradients, mode collapse, and a lack of stopping criteria, along with the significant computational demands, push them toward the limits of current classical computers. Recent theoretical studies suggest that quantum algorithms have the potential to outperform their classical counterparts in a variety of tasks \cite{nielsen}, \cite{Montanaro_2016}, \cite{arute2019quantum}. On the other hand, Zoufal et.al.\cite{zoufal} showed that qGANs can efficiently learn and load generic probability distributions into quantum states, making qGANs a promising method in the search for alternative future HEP simulation approaches. 

\paragraph{Generative Adversarial Networks} is a framework for estimating generative models using an adversarial process \cite{goodfellow}. This framework involves training two models simultaneously: a generative model \textbf{G} that captures the data distribution and a discriminative model \textbf{D} that distinguishes between samples from the training data distribution and those produced by \textbf{G}. The goal is to improve \textbf{G} so that \textbf{D} cannot differentiate between training data and generated data. This process is similar to a min-max two-player game, where \textbf{G} tries to fool \textbf{D} while \textbf{D} aims to detect the fake data. We train \textbf{D} to maximize the probability of assigning the correct label to both the training data samples and the generated samples from \textbf{G}. Simultaneously, we train \textbf{G} to minimize the chances of \textbf{D} correctly distinguishing between training and generated samples. This process involves optimizing the loss function \cite{goodfellow}:

\begin{equation*}
\begin{aligned}
    V(D, G) &= \mathbb{E}_{\mathbf{x} \sim p_{\text{data}}(\mathbf{x})} [\log D(\mathbf{x})]  + \mathbb{E}_{\mathbf{z} \sim p_{\mathbf{z}}(\mathbf{z})} [\log (1 - D(G(\mathbf{z})))]
\end{aligned}
\end{equation*}

where $D(\mathbf{x})$ represents the probability that $\mathbf{x}$ came from the true data and $D(G(\mathbf{z}))$ the probability of $D$ correctly labeling a generated sample $G(\mathbf{z})$ from the latent space.

\paragraph{Quantum Generative Adversarial Networks} represent a quantum extension of classical GANs, incorporating quantum mechanics to leverage the computational advantages of quantum systems \cite{dallaire}, \cite{lloyd}. In qGANs, the generator is a parameterized quantum circuit (PQC) that produces quantum states resembling the distribution from the training data, meanwhile, the discriminator can be either a classical discriminator or a PQC that differentiates between the training data distribution and generated distribution. When quantum generators produce high-dimensional measurement statistics, the qGAN may demonstrate quantum advantages over classical GANs. As a result, qGANs can achieve faster convergence or utilize fewer physical resources \cite{Tong}.

In this work we train and evaluate our proposed (qGAN) on multi-detector jet images obtained from the CMS Open Data Portal \cite{cern_opendata}. The qGAN's objective is to accurately generate jet images, a critical task for improving the efficiency of simulations and analysis in high-energy physics experiments at the Large Hadron Collider (LHC).

\section{Method}
\subsection{Data}
\begin{figure}[ht]
    \centering    
    \includegraphics[width=\linewidth]{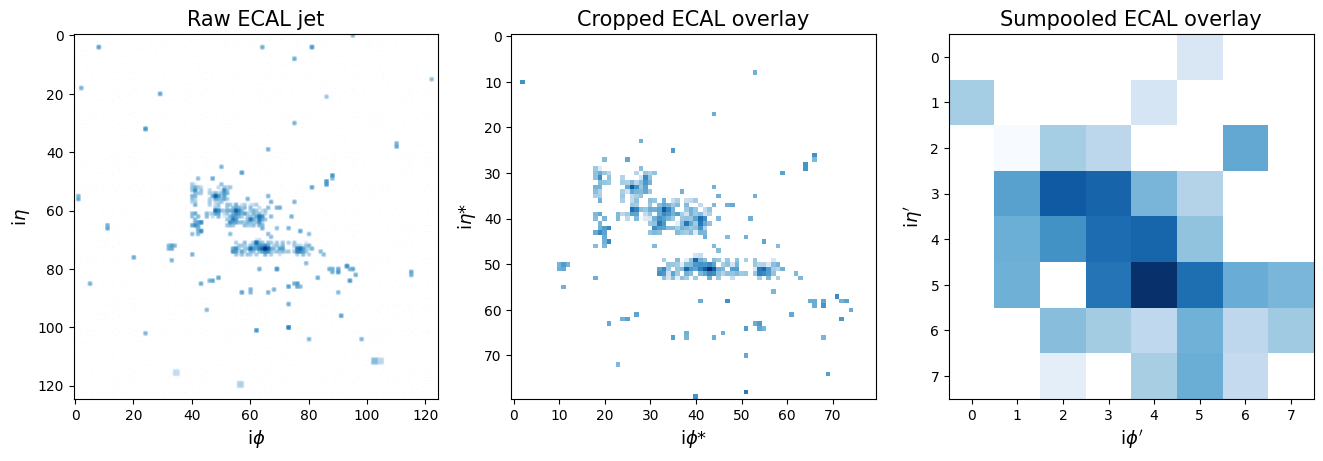}
    \caption{Data preprocessing steps for ECAL jet images. The left panel shows the raw ECAL jet with a resolution of 125x125 pixels. The middle panel depicts the cropped ECAL overlay, focusing on the central region with a reduced resolution of 80x80 pixels. The right panel illustrates the sum-pooled ECAL overlay, further downsampled to an 8x8 resolution. All plotted in a log scale.}
    \label{fig: preprocessing}
\end{figure}

We utilize the dataset described by Andrews et al. \cite{Andrews_2020}, generated from simulated data for QCD dijet production available on the CERN CMS Open Data Portal \cite{cern_opendata}. The dataset includes 933,206 three-channel images, each measuring 125 × 125 pixels, with half representing quarks and the other half representing gluons. Each channel in the images corresponds to a specific component of the Compact Muon Solenoid (CMS) detector: the inner tracking system (Tracks), which identifies charged particle tracks; the electromagnetic calorimeter (ECAL), which records energy deposits from electromagnetic particles; and the hadronic calorimeter (HCAL), which captures energy deposits from hadrons. In this work, we used the ECAL and HCAL channels from gluon images to train and evaluate the model. Due to computational constraints, we pre-process gluon images before training. As the relevant information is in the center of the pictures, these were cropped to $80\times 80$ pixel images. This cropping led to a slight reduction in the total energy deposited, as we excluded the peripheral pixels. Next, we down-scaled the cropped images using sum pooling. This process applies a $10\times 10$ kernel with a stride of ten and no padding, summing the pixel values within each kernel region. The final result of the pre-processing consist of images of an 8x8 resolution representing the scaled energy deposits in the ECAL and HCAL channels. The preprocessing steps are shown in the figure \ref{fig: preprocessing}.

The preprocessing steps applied to the jet images ensure that the essential features of the original data are preserved while reducing computational complexity. Cropping the raw images from 125x125 to 80x80 pixels focuses on the central region of the jets, where the majority of the energy is deposited. This process removes peripheral pixels with negligible energy contributions, retaining the critical energy distribution patterns essential for jet analysis. Subsequently, sum pooling further reduces the image resolution from 80x80 to 8x8 pixels by summing the pixel intensities within 10x10 kernels. This step preserves the overall energy distribution and spatial correlations, with each pixel in the 8x8 images representing the localized energy sum of a specific region. By focusing on the central high-energy regions and maintaining the global and local energy patterns, these preprocessing steps ensure that the physics-relevant characteristics of the jets remain intact.

\subsection{Architecture}
\begin{figure}[ht]
    \centering
    \begin{subfigure}[b]{0.35\textwidth} 
        \centering        
        \includegraphics[width=\linewidth]{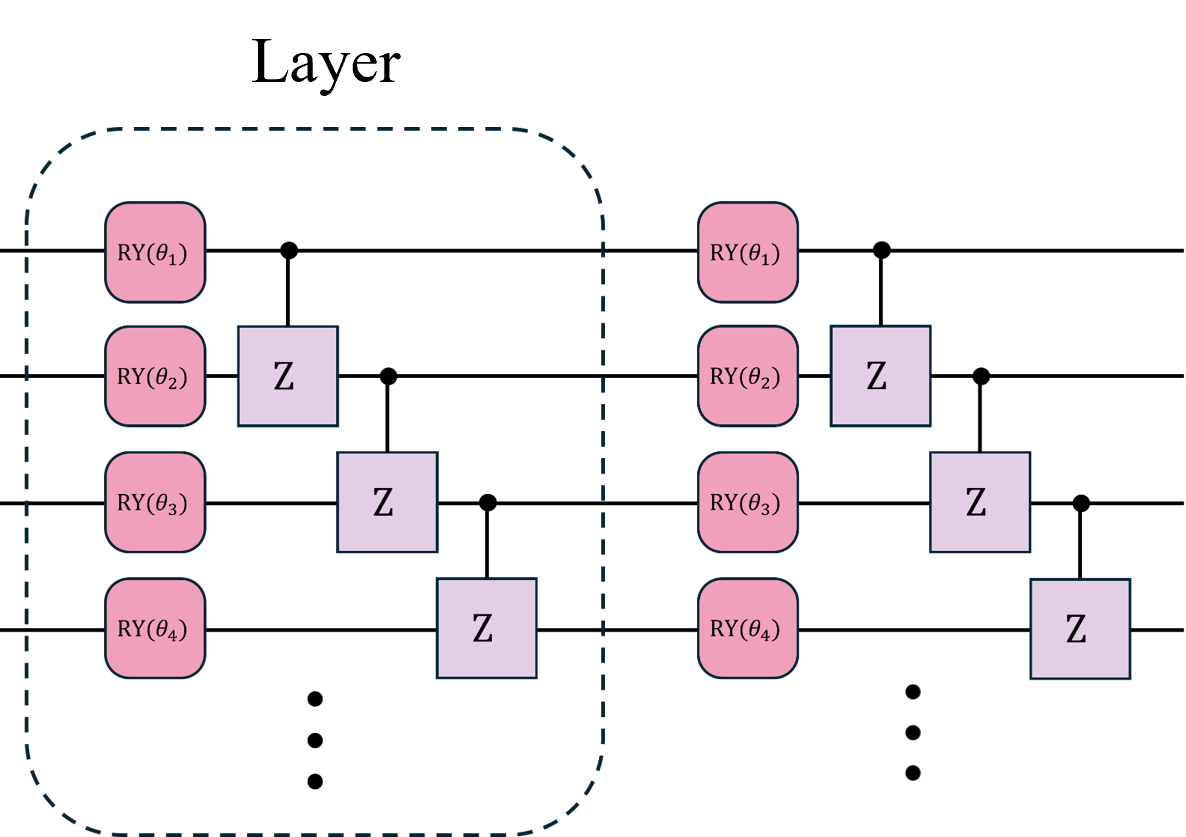}
    \end{subfigure}
    \hfill
    \begin{subfigure}[b]{0.60\textwidth} 
        \centering        
        \includegraphics[width=\linewidth]{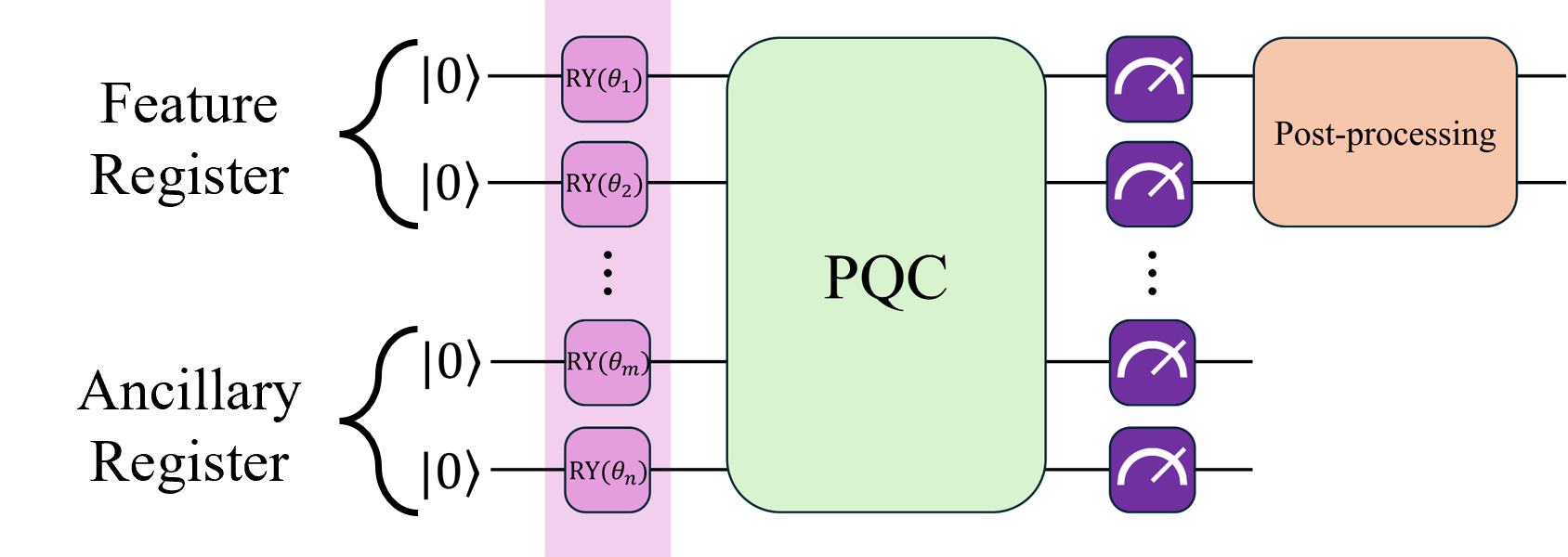}
    \end{subfigure}
    \caption{Schematic representation of the quantum generator. The PQC (left) includes both a feature register and an ancillary register along with a post-processing step after measurement. A single layer (right) consisting of Pauli-Y rotations followed by a control Z gate entangling block.}
    \label{fig: parameterized circuit}
\end{figure}

In this work, we use the hybrid architecture proposed by Huang et. al. \cite{Huang_2021}, composed of many generators, each responsible for producing a fraction of the image. This approach allows each generator to focus on a simpler distribution rather than the more complex complete image and a classical discriminator. The generated images are constructed using the probabilities of quantum basis states, where each basis state probability directly corresponds to the intensity of a pixel in the final image. We use PQCs as generators with reference state $|\Psi_{in}\rangle$ prepared according to a discrete uniform distribution as followed by layers of parameterized Pauli-Y rotations along with an entangling block of control Z gates. This architecture uses an auxiliary register to provide the PQC with more possible solution states through a non-linear transformation. The individual layers and the complete PQC are shown in figure \ref{fig: parameterized circuit}. 

On the other hand, the discriminator is a dense neural network consisting of a 64-node input layer, two hidden layers of 128 and 32 nodes respectively both followed by ReLu activation function. Finally, the output layer consist of a single neuron, applying a Sigmoid activation function. The choice of a hybrid quantum-classical architecture over purely quantum architecture stems from its ability to leverage the strengths of both paradigms. Quantum generators can explore complex data distributions more efficiently due to quantum entanglement and superposition, providing potential exponential speedups in sampling. Meanwhile, classical discriminators are well-suited for large-scale classification tasks. This hybrid approach balances quantum advantages, such as faster convergence and reduced resource requirements, with the robustness and scalability of classical models.

\subsubsection{Non-linear transform}
Since quantum gates are inherently linear transformations, more complex data distributions may require non-linear transformations for more accurate results. For the pre-measurement state of the generator, we have:

\begin{equation*}
    |\Psi(z)\rangle = U_G(\theta)|z\rangle
\end{equation*}

Where $U_G(\theta)$ represents the overall unitary operator of the parametrized layers. If we take a partial measurement $\Pi$ and trace out the ancillary subsystem:

\begin{equation}
\begin{aligned}
    V(D, G) &= \frac{\mathrm{Tr}_A (\Pi \otimes |\Psi(z)\rangle \langle \Psi(z)|)}{\mathrm{Tr} (\Pi \otimes |\Psi(z)\rangle \langle \Psi(z)|)} = \frac{\mathrm{Tr}_A (\Pi \otimes |\Psi(z)\rangle \langle \Psi(z)|)}{\langle \Psi(z) | \Pi \otimes |\Psi(z)\rangle}
\label{non-linear_transformation}
\end{aligned}
\end{equation}

The post-measurement state $p(z)$, is dependent on $z$ on the numerator and denominator. This implies a non-linear transformation was performed over $|\Psi(z)\rangle$.

\subsubsection{Post-processing}
Along with the non-linear transformation we performed a post-processing step that allows the output of the quantum generator to get rid of the normalization constraint of the measurement. As we want the output of the quantum generator to represent energy deposits, the output should be able to take values larger than 1, and the elements from the quantum circuit output do not necessarily need to sum up to 1. These limitations can be overcome by applying the following transformation to the quantum circuit output of the circuit.

\begin{equation}
    \tilde{x} = \frac{\textbf{g}}{y}
    \label{output transformation}
\end{equation}

Where $\textbf{g}$ is the vector with the basis states probabilities and $y \in (0, 1)$. 

\subsection{Experiments}
After generating the quantum states representing the jet images, a partial measurement is performed to condition the results on auxiliary qubits. We simulate outcome probabilities and apply a non-linear transformation \ref{non-linear_transformation}. The post-processing step adjusts the probabilities by scaling them with a factor $y$ from eq.\ref{output transformation} and filtering out values below a threshold of 0.001 to remove negligible values. This ensures that the final values reflect realistic energy distributions. 
We use a training dataset of 512 gluon-initiated images due to computational constraints and a batch size of 1 for faster convergence. The generator consists of 4 PQCs of 5 feature qubits, 2 auxiliary qubits, and a depth of 10 layers. We train the model for 50 epochs using a stochastic gradient descent optimizer and learning rates of 0.001 and 0.005 for the generator and discriminator respectively. The hyperparameter's initialization follows a uniform distribution in the interval $[0, 1)$ and the value of $y$ in the output transformation \ref{output transformation} is 0.28. We use pennylane and torch frameworks and run on a machine with AMD Ryzen 5 5600G CPU and 16GB RAM resources, the training time with these resources is about 300 minutes. To ensure reproducibility, we use a fixed random seed, set to 42.

\section{Results}
\begin{figure}[ht]
    \centering
    \begin{subfigure}[b]{0.45\textwidth}
         \centering
         \includegraphics[width=\linewidth]{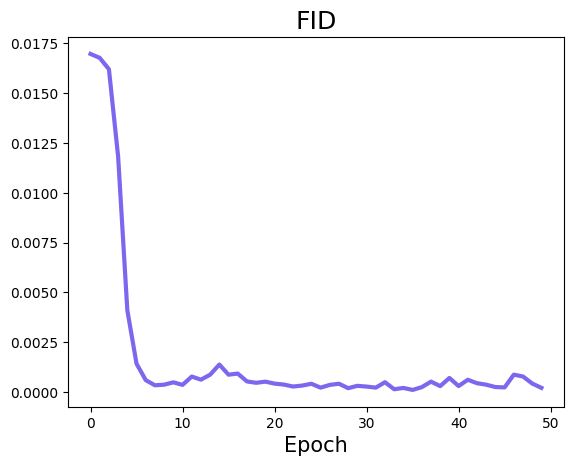}
    \end{subfigure}
    \hfill
    \begin{subfigure}[b]{0.44\textwidth}
         \centering
         \includegraphics[width=\linewidth]{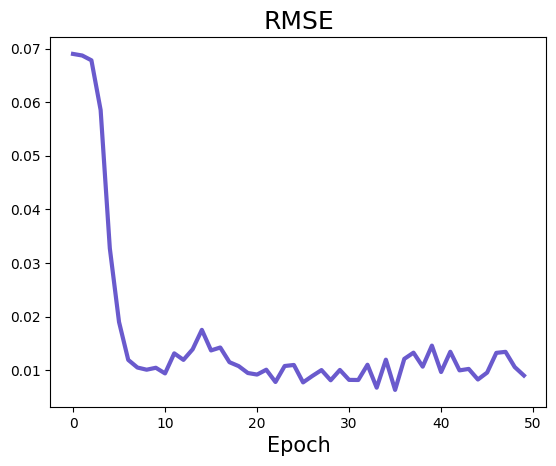}
    \end{subfigure}
    \caption{Frenchet inception distance (left) and the root mean squared error (right) computed using the training data overlay and the generated data overlay at the end of each epoch.}
    \label{fig: training metrics}
\end{figure}

For the evaluation of the generative model \cite{Kansal_2023}, The evolution of the Frechet Inception Distance (upper plot) and the Root Mean Squared Error (lower plot) during training is shown in the figure \ref{fig: training metrics}. 

\begin{figure}[ht]
    \centering
    \begin{subfigure}[b]{0.48\textwidth} 
         \centering
         \includegraphics[width=\linewidth]{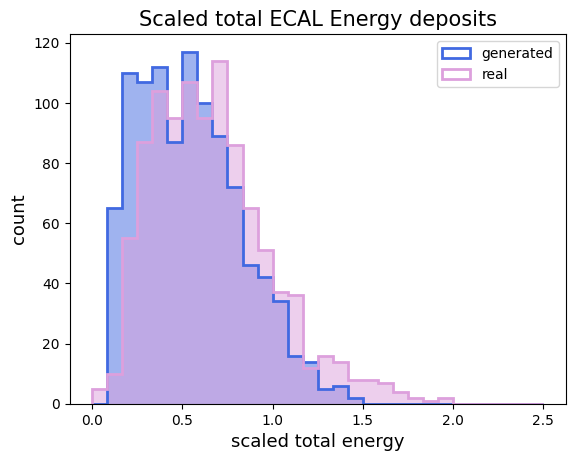}
    \end{subfigure}
    \hfill
    \begin{subfigure}[b]{0.48\textwidth}
         \centering
         \includegraphics[width=\linewidth]{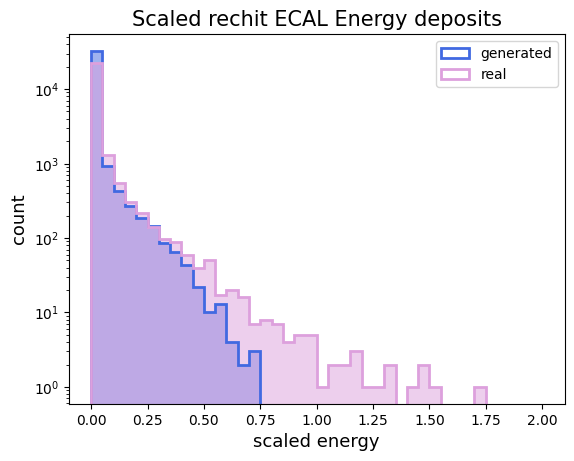}
    \end{subfigure}
    \caption{Comparison of the generated and real ECAL energy deposits. The left plot shows the scaled total ECAL energy deposits. The right plot depicts the scaled rechit ECAL energy deposits plotted in a log scale.}
    \label{fig: scaled energies ECAL}
\end{figure}

\begin{figure}[ht]
    \centering
    \begin{subfigure}[b]{0.48\textwidth}
         \centering
         \includegraphics[width=\linewidth]{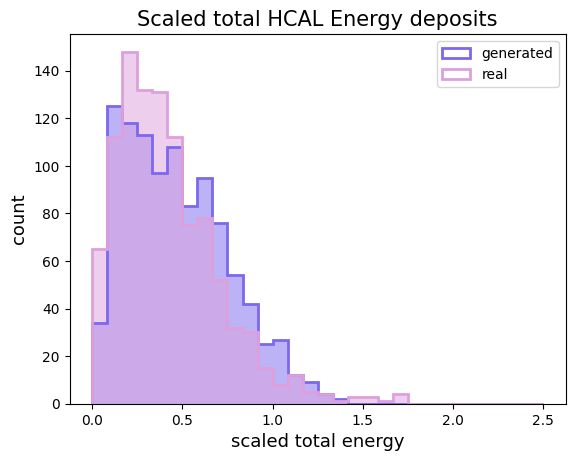}
    \end{subfigure}
    \begin{subfigure}[b]{0.48\textwidth}
         \centering
         \includegraphics[width=\linewidth]{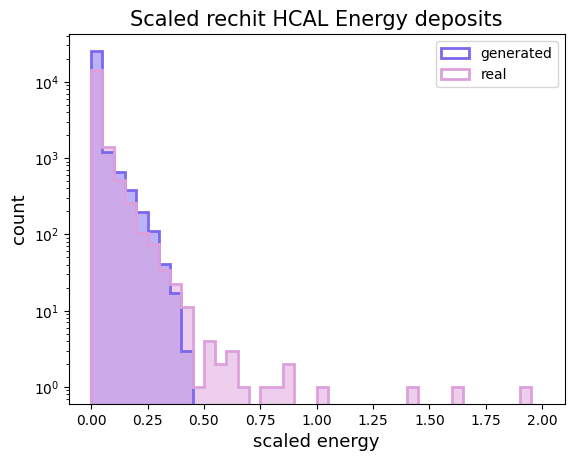}
    \end{subfigure}
    \caption{Comparison of the generated and real HCAL energy deposits. The left plot shows the scaled total HCAL energy deposits. The right plot depicts the scaled rechit HCAL energy deposits plotted in a log scale.}
    \label{fig: scaled energies HCAL}
\end{figure}

\begin{figure}[ht]
    \centering    
    \includegraphics[width=0.48\textwidth]{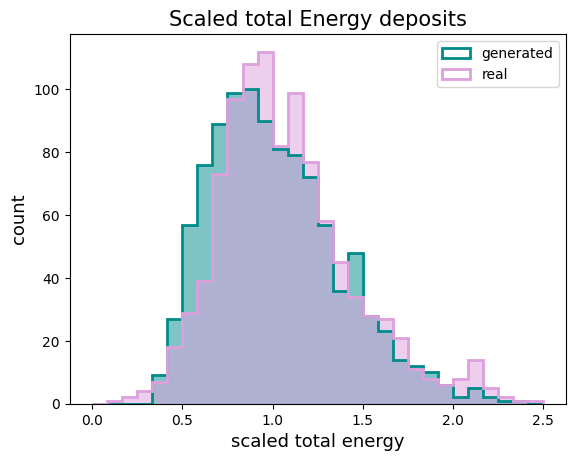}
    \caption{Comparison of the scaled total energy deposits for real and generated jets. The histogram shows the distribution of total energy deposits, calculated as the sum of deposited energy in the ECAL and HCAL channels, for real and generated jets. }
    \label{fig:scaled total energy}
\end{figure}

We observe a clear convergence and a sharp initial decrease in both of them, this indicates the model's fast convergence and increased accuracy in generating realistic gluon images.
To evaluate if the model learned the implicit features of the training dataset, the scaled total energy deposited by a single jet in the ECAL and HCAL channels is analyzed. Additionally, the rechit energy deposits, which correspond to the energy deposited by a single particle from a jet, are compared against the same distributions from the generated gluon jets. As shown in figures \ref{fig: scaled energies ECAL} and \ref{fig: scaled energies HCAL} it is interesting to note that both distributions from ECAL and HCAL channels were concurrently generated with a high degree of accuracy, The final evaluation, shown in Figure \ref{fig:scaled total energy}, compares the total energy of real and generated jets. This total energy is calculated as the sum of the energies from the ECAL and HCAL channels. This comparison evaluates the model's ability to learn the implicit correlation between the channels for the same event. The results demonstrate that the model successfully captures and reproduces the implicit features present in the training data.

\begin{figure}[ht]
    \centering
    \begin{subfigure}[b]{0.48\columnwidth}
         \centering
         \includegraphics[width=\linewidth]{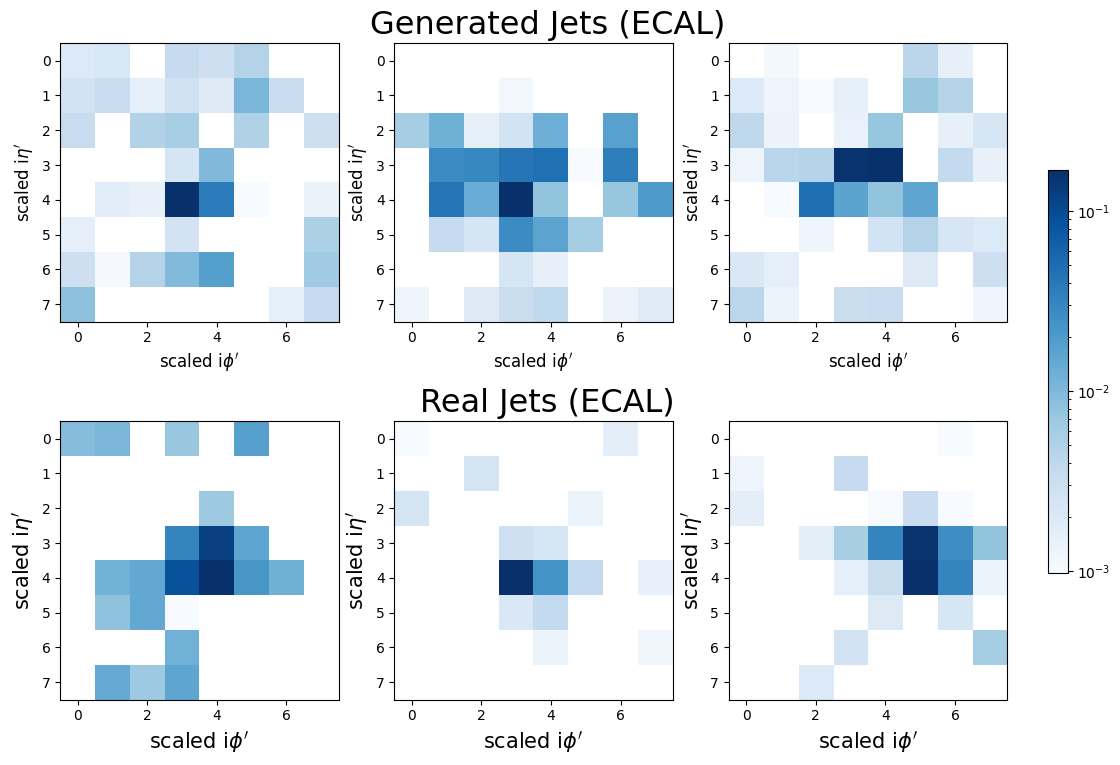}
    \end{subfigure}
    \hfill
    \begin{subfigure}[b]{0.48\columnwidth}
         \centering
         \includegraphics[width=\linewidth]{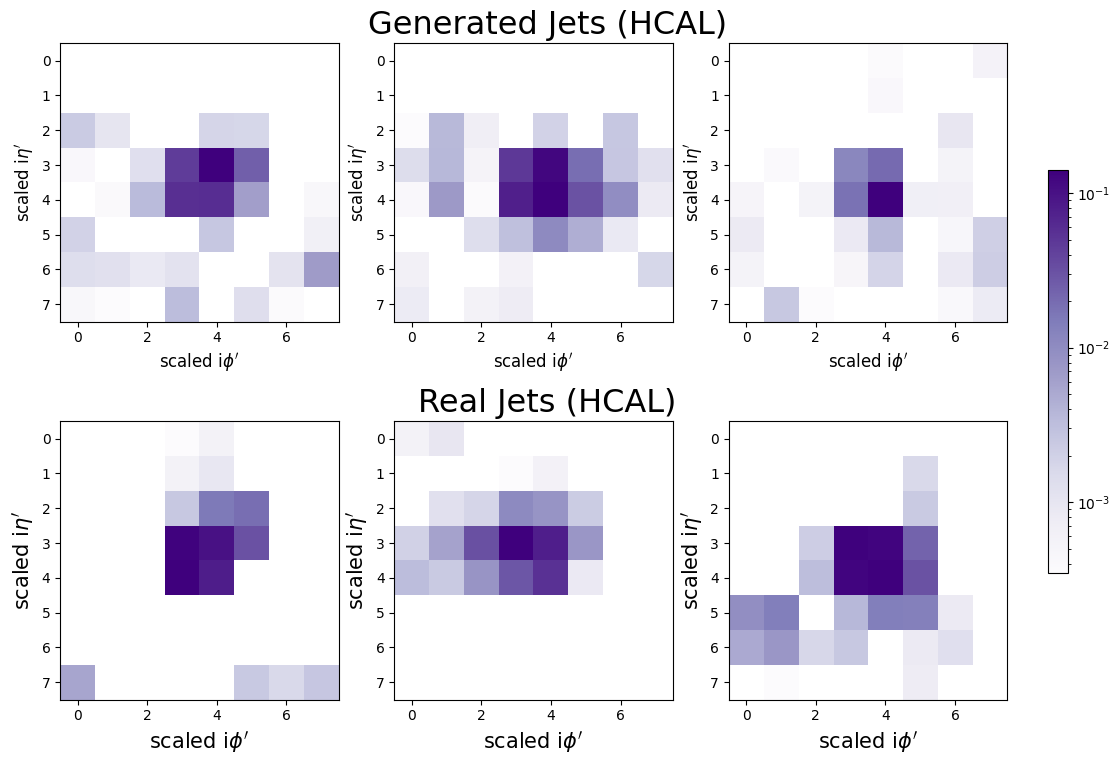}
    \end{subfigure}
    \caption{Comparison of real and generated jet energy distributions in ECAL and HCAL channels. The left plot displays the scaled energy deposits for three examples of generated jets (top row) and three examples of real jets (second row) in the ECAL. The right plot shows analogous comparisons for the HCAL, with the first row depicting three examples of generated jets and the bottom row showing three examples of real jets. Color intensity represents the magnitude of the energy deposited. Both sets are plotted in a log scale.}
    \label{fig: individual jets}
\end{figure}

Figure \ref{fig: individual jets} compares real and generated ECAL and HCAL jet images. The upper panel presents three examples of real ECAL jets and three generated ECAL jets from the trained qGAN model for comparison. Similarly, the lower panel showcases three examples of real HCAL jets and three generated HCAL jets from the same model. Both sets exhibit similar energy deposition patterns, with the generated jets closely resembling the real jets regarding core intensity and distribution of the energy deposits.

\begin{figure}[ht]
    \centering
    \begin{subfigure}[b]{0.6\columnwidth}
         \centering
         \includegraphics[width=\linewidth]{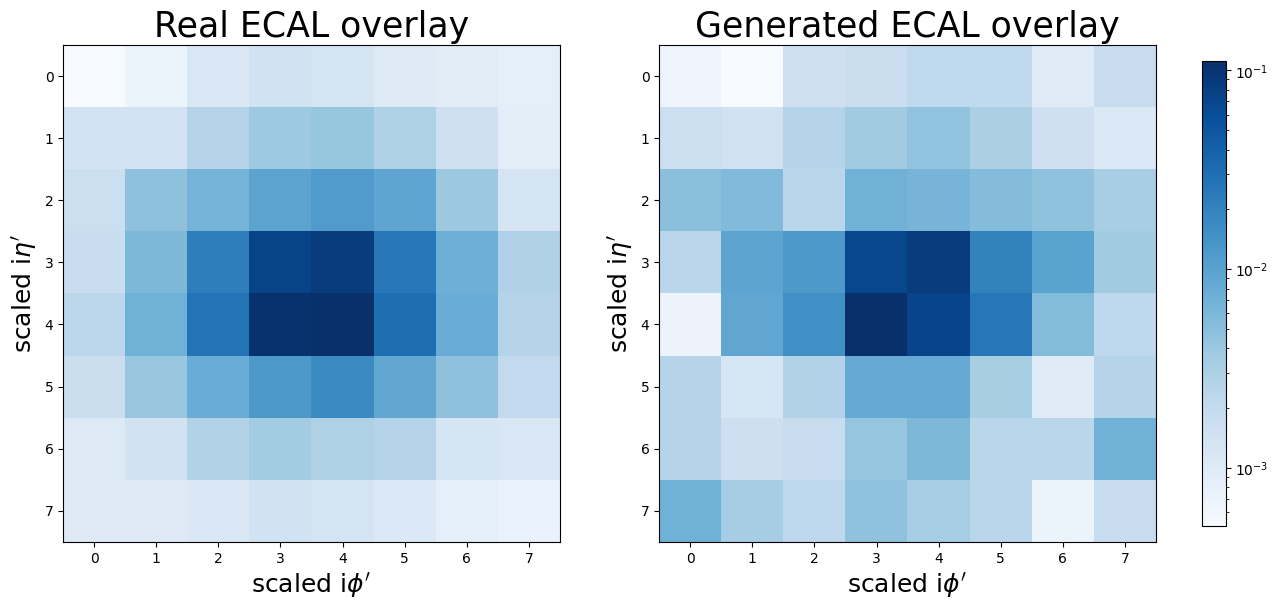}
    \end{subfigure}
    \begin{subfigure}[b]{0.6\columnwidth}
         \centering
         \includegraphics[width=\linewidth]{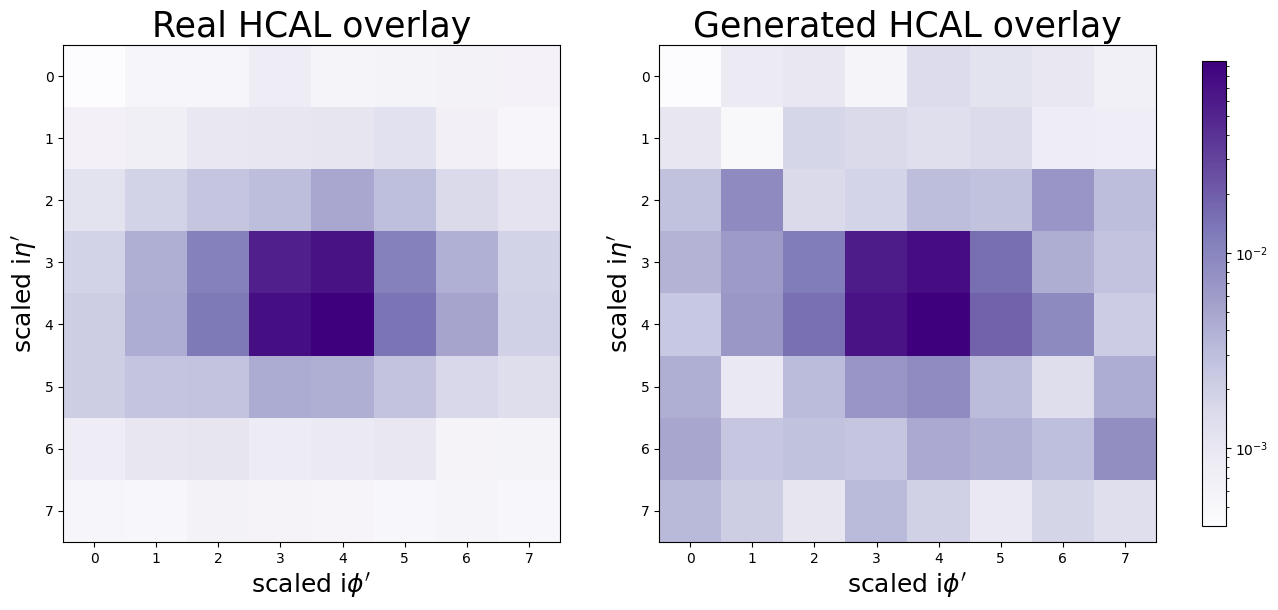}
    \end{subfigure}
    \caption{Real and generated ECAL and HCAL channels' jet-view image overlays. In the upper plots, the left panel shows the real ECAL overlay, while the right panel shows the generated ECAL overlay over 512 jets each, In the bottom plots the analogous comparisons for the HCAL, where the left panel shows the real HCAL overlay, while the right panel shows the generated HCAL overlay over 512 jets each. All plotted in log-scale. Image resolution $8 \times 8$.}
    \label{fig: overlays}
\end{figure}

Finally, Figure \ref{fig: overlays} presents a comparison of the energy deposit overlays for both the ECAL and HCAL channels, highlighting the similarities between real and generated jets. For the ECAL channel, the real energy deposit overlay (top left) and the generated energy deposit overlay (top right) show comparable energy distributions, with both centered around the high-energy deposit region. The generated overlay closely replicates the real data, particularly in the core region, demonstrating the qGAN's ability to capture the spatial distribution of energy deposits in the ECAL channel. Similarly, for the HCAL channel, the real overlay (bottom left) and generated overlay (bottom right) exhibit similar patterns, with the generated overlay effectively reproducing the key features of the real energy distributions. These results confirm the model's capability to learn and replicate the energy deposition patterns in both calorimeter channels.

\section{Conclusions}
In this work, we successfully implemented a qGAN to generate ECAL and HCAL sub-detectors gluon-initiated jet images from the CMS Open Data simultaneously. The results demonstrate the ability of qGANs to learn and replicate energy deposit distributions with high accuracy, marking a significant step in advancing quantum approaches in high-energy physics simulations. However, the current work was limited to 512 images of a downscaled resolution. Additionally, training was performed on quantum simulators, which may not fully capture the behavior on actual quantum hardware. These limitations could affect the scalability and generalization of the qGAN, highlighting areas for future work. Exploring real quantum hardware, expanding the dataset, and integrating data from additional sub-detectors will be key next steps to enhance the model’s robustness and applicability. This study sets the foundation for future work to improve the qGAN model to simultaneously generate three-channel images and incorporate quark jet image generation, enhancing the simulation capabilities for LHC-related tasks.

\section{Data Availability}

The code is open source and can be found in the following repository: \href{https://github.com/ReyGuadarrama/QGAN_for_MC_Simulations}{https://github.com/ReyGuadarrama/QGAN\_for\_MC\_Simulations}

\clearpage

\bibliographystyle{unsrt}  
\bibliography{references}

\begin{thebibliography}{10}

\bibitem{metropolis}
N.~Metropolis, A.W. Rosenbluth, M.N. Rosenbluth, A.H. Teller, and E.~Teller.
\newblock {Equation of State Calculations by Fast Computing Machines}.
\newblock {\em Journal of Chemical Physics}, 21(6):1087--1092, 1953.

\bibitem{Carlson}
J.~Carlson, S.~Gandolfi, F.~Pederiva, Steven~C. Pieper, R.~Schiavilla, K.~E. Schmidt, and R.~B. Wiringa.
\newblock Quantum monte carlo methods for nuclear physics.
\newblock {\em Rev. Mod. Phys.}, 87:1067--1118, Sep 2015.

\bibitem{Colin}
Colin Morningstar.
\newblock The monte carlo method in quantum field theory, 2007.

\bibitem{goodfellow}
Ian~J. Goodfellow, Jean Pouget-Abadie, Mehdi Mirza, Bing Xu, David Warde-Farley, Sherjil Ozair, Aaron Courville, and Yoshua Bengio.
\newblock Generative adversarial networks, 2014.

\bibitem{goodfellow2016tutorial}
Ian Goodfellow.
\newblock Nips 2016 tutorial: Generative adversarial networks, 2017.

\bibitem{Musella_2018}
Pasquale Musella and Francesco Pandolfi.
\newblock Fast and accurate simulation of particle detectors using generative adversarial networks.
\newblock {\em Computing and Software for Big Science}, 2(1), November 2018.

\bibitem{Di_Sipio_2019}
Riccardo Di~Sipio, Michele~Faucci Giannelli, Sana~Ketabchi Haghighat, and Serena Palazzo.
\newblock Dijetgan: a generative-adversarial network approach for the simulation of qcd dijet events at the lhc.
\newblock {\em Journal of High Energy Physics}, 2019(8), August 2019.

\bibitem{nielsen}
Michael~A. Nielsen and Isaac~L. Chuang.
\newblock {\em {Quantum Computation and Quantum Information}}.
\newblock Cambridge University Press, 2010.

\bibitem{Montanaro_2016}
Ashley Montanaro.
\newblock Quantum algorithms: an overview.
\newblock {\em npj Quantum Information}, 2(1), January 2016.

\bibitem{arute2019quantum}
Frank Arute, Kunal Arya, Ryan Babbush, et~al.
\newblock Quantum supremacy using a programmable superconducting processor.
\newblock {\em Nature}, 574(7779):505--510, 2019.

\bibitem{zoufal}
C.~Zoufal, A.~Lucchi, and S.~Woerner.
\newblock {Quantum Generative Adversarial Networks for learning and loading random distributions}.
\newblock {\em npj Quantum Information}, 5(1):103, 2019.

\bibitem{dallaire}
Pierre-Luc Dallaire-Demers and Nathan Killoran.
\newblock Quantum generative adversarial networks.
\newblock {\em Phys. Rev. A}, 98:012324, Jul 2018.

\bibitem{lloyd}
Seth Lloyd and Christian Weedbrook.
\newblock Quantum generative adversarial learning.
\newblock {\em Phys. Rev. Lett.}, 121:040502, Jul 2018.

\bibitem{Tong}
Tong Li, Shibin Zhang, and Jinyue Xia.
\newblock Quantum generative adversarial network: A survey.
\newblock {\em Computers, Materials \& Continua}, 64(1):401--438, 2020.

\bibitem{cern_opendata}
{CERN Open Data Portal}.
\newblock {CMS Open Data Information}.
\newblock \url{https://opendata.cern.ch/docs/about-cms}, 2024.

\bibitem{Andrews_2020}
M.~Andrews, J.~Alison, S.~An, B.~Burkle, S.~Gleyzer, M.~Narain, M.~Paulini, B.~Poczos, and E.~Usai.
\newblock End-to-end jet classification of quarks and gluons with the cms open data.
\newblock {\em Nuclear Instruments and Methods in Physics Research Section A: Accelerators, Spectrometers, Detectors and Associated Equipment}, 977:164304, October 2020.

\bibitem{Huang_2021}
He-Liang Huang, Yuxuan Du, Ming Gong, Youwei Zhao, Yulin Wu, Chaoyue Wang, Shaowei Li, Futian Liang, Jin Lin, Yu~Xu, Rui Yang, Tongliang Liu, Min-Hsiu Hsieh, Hui Deng, Hao Rong, Cheng-Zhi Peng, Chao-Yang Lu, Yu-Ao Chen, Dacheng Tao, Xiaobo Zhu, and Jian-Wei Pan.
\newblock Experimental quantum generative adversarial networks for image generation.
\newblock {\em Physical Review Applied}, 16(2), August 2021.

\bibitem{Kansal_2023}
Raghav Kansal, Anni Li, Javier Duarte, Nadezda Chernyavskaya, Maurizio Pierini, Breno Orzari, and Thiago Tomei.
\newblock Evaluating generative models in high energy physics.
\newblock {\em Physical Review D}, 107(7), April 2023.

\end{thebibliography}

\end{document}